\documentstyle[pre,aps,epsf,floats]{revtex}

\begin{document}

\draft

\title{Multi-scaling properties of truncated L\'evy flights}

\author{Hiroya Nakao}

\address{Graduate School of Mathematical Sciences, University of
  Tokyo,\\ 3-8-1 Komaba, Meguro, Tokyo 153-8914, Japan}

\date{\today}

\maketitle

\begin{abstract}
  Multi-scaling properties of one-dimensional truncated L\'evy flights
  are studied.
  Due to the broken self-similarity of the distribution of jumps, they
  are expected to possess multi-scaling properties in contrast to the
  ordinary L\'evy flights.
  We argue this fact based on a smoothly truncated L\'evy
  distribution, and derive the functional form of the scaling
  exponents.
  Specifically, they exhibit bi-fractal behavior, which is the
  simplest case of multi-scaling.
\end{abstract}

\section{Introduction}

L\'evy flights are random processes based on L\'evy stable
distributions\cite{Feller}. They have been utilized successfully in
modeling various complex spatio-temporal behavior of non-equilibrium
dissipative systems such as fluid turbulence, anomalous diffusion, and
financial markets\cite{Shlesinger}.

The L\'evy stable distribution is self-similar to its convolutions.
It has a long power-law tail that decays much more slowly compared
with an exponential decay, which gives rise to infinite variance.
However, in practical situations, it is usually truncated due to
nonlinearity or finiteness of the system.
In order to incorporate this fact, Mantegna and Stanley\cite{Mantegna}
introduced the notion of truncated L\'evy flight.
It is based on a truncated L\'evy stable distribution with a sharp
cut-off in its power-law tail. Therefore, the distribution is not
self-similar when convoluted, and has finite variance.
Thus, the truncated L\'evy flight converges to a Gaussian process due
to the central limit theorem in contrast to the ordinary L\'evy stable
process.
However, as they pointed out, its convergence to a Gaussian is very
slow and the process exhibits anomalous behavior in a wide range
before the convergence.
Koponen\cite{Koponen} reproduced their result analytically using a
different type of truncated L\'evy distribution with a smoother
cut-off.

Dubrulle and Laval\cite{Dubrulle} applied their idea to the velocity
field of 2D fluid turbulence, and claimed that the truncation is
essential for the multi-scaling property of the velocity field to
appear.
They showed that the broken self-similarity of the distribution makes
a qualitative difference on the scaling property of the corresponding
random process, i.e., the ordinary L\'evy stable process exhibits mere
single-scaling, while the truncated L\'evy process exhibits
multi-scaling, although their analysis was mostly based on numerical
calculation and the obtained scaling exponents were rather inaccurate.
The idea of truncated L\'evy flights has also been applied to the
analysis of financial data such as stock market prices and foreign
exchange rates\cite{Cont,Bouchaud}. Multi-scaling analyses of the
financial data have also been attempted\cite{Vandewalle,Ivanova}.

In this paper, we treat the truncated L\'evy flights analytically
based on the smooth truncation introduced by Koponen, and clarify
their multi-scaling properties.
They exhibit the simplest form of multi-scaling, i.e., bi-fractality,
due to the characteristic shape of the truncated L\'evy distribution.
Our results may have some relevance to the multi-scaling properties of
the velocity field in 2D fluid turbulence and of the fluctuation of
the stock market prices.

\begin{figure}[htbp]
  \begin{center}
    \leavevmode
    \epsfxsize=0.5\textwidth
    \epsfbox{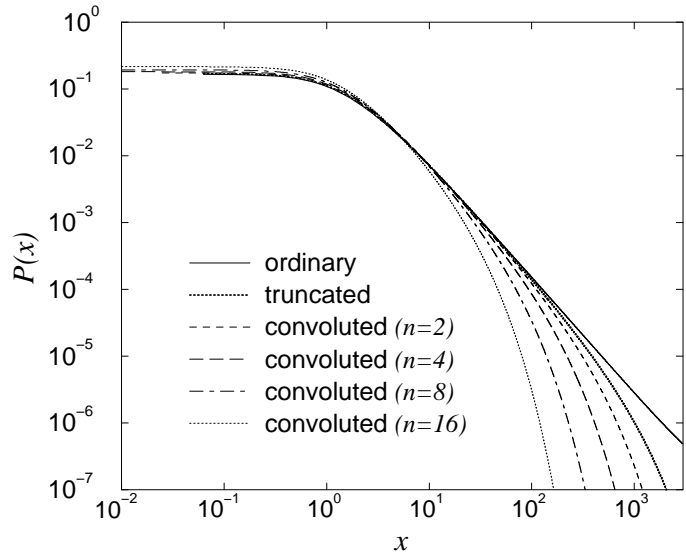}
    \caption{Truncated L\'evy distribution $P_{TL}(x ; \lambda)$ 
      and its $n$-times convolutions for the symmetric case. The
      parameters are $\lambda=0.001$, $\alpha=0.75$, and
      $q-p=0$. Corresponding ordinary L\'evy stable distribution
      $P_{L}(x)$ is also shown for comparison. Each convoluted
      distribution is rescaled as $n^{-1/\alpha} P(n^{-1/\alpha} x)$.}
    \label{Fig:01}
  \end{center}
\end{figure}

\begin{figure}[htbp]
  \begin{center}
    \leavevmode
    \epsfxsize=0.5\textwidth
    \epsfbox{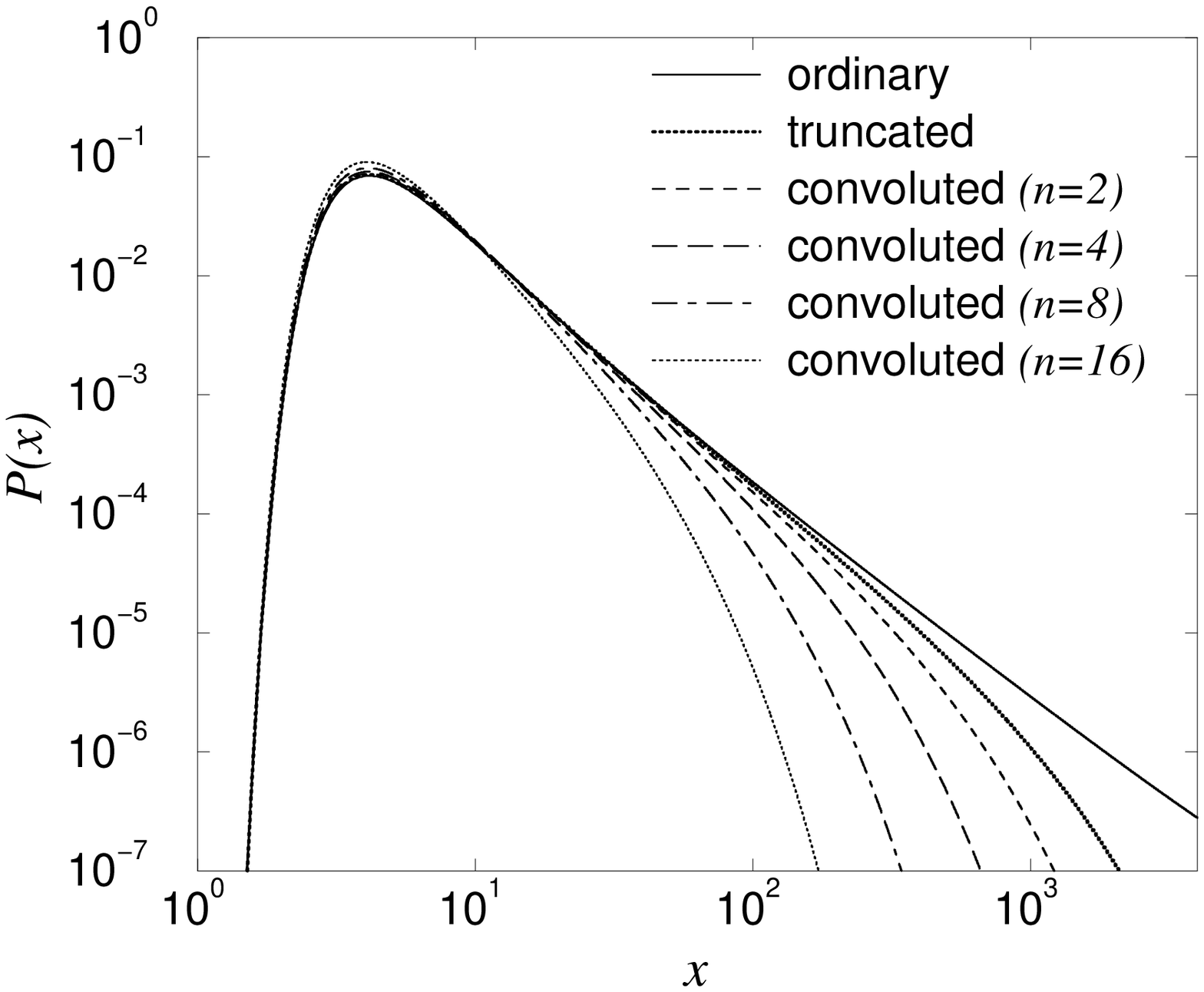}
    \caption{Truncated L\'evy distribution $P_{TL}(x ; \lambda)$ 
      and its $n$-times convolutions for the one-sided case. The
      parameters are $\lambda=0.001$, $\alpha=0.75$, and
      $q-p=-1$. Corresponding ordinary L\'evy stable distribution
      $P_{L}(x)$ is also shown for comparison. Each convoluted
      distribution is rescaled as $n^{-1/\alpha} P(n^{-1/\alpha} x)$.}
    \label{Fig:02}
  \end{center}
\end{figure}

\section{Truncated L\'evy distribution}

Let $P(x)$ an probability distribution and $e^{\psi(\zeta)}$ its
characteristic function, i.e.,
\begin{equation}
  P(x) = \frac{1}{2 \pi} \int_{-\infty}^{\infty} e^{\psi(\zeta)} e^{-i
    \zeta x} d\zeta, \;\;\;\; e^{\psi(\zeta)} =
  \int_{-\infty}^{\infty} P(x) e^{i \zeta x} dx.
  \label{Eq:Fourier}
\end{equation}
As explained in Feller\cite{Feller}, (the argument of) the
characteristic function for a L\'evy stable distribution is given by
that of a compound Poisson process:
\begin{equation}
  \psi(\zeta) = \int_{-\infty}^{\infty} \left( e^{i \zeta x} - 1 - i
    \zeta \tau(x) \right) f(x) \ dx,
\end{equation}
where $f(x)$ is a probability distribution of increments and $\tau(x)$
is a certain centering function.
$f(x)$ is assumed to be
\begin{equation}
  f_{L}(x) = \left\{
    \begin{array}{ll}
      C q \ |x|^{-1-\alpha} \;\;&\;\; (x<0), \cr
      C p \ x^{-1-\alpha} \;\;&\;\; (x>0), \cr
    \end{array}
  \right.
  \label{Eq:increments}
\end{equation}
where $C > 0$ is a scale constant, $0 < \alpha < 2$, $p \geq 0$, $q
\geq 0$, and $p+q=1$. The function $\tau(x)$ is chosen as $0$ for $0 <
\alpha < 1$ and $x$ for $1 < \alpha < 2$.
By integration, we obtain
\begin{equation}
  \psi_L(\zeta) = C \Gamma(-\alpha) |\zeta|^{\alpha} \left( \cos
    \frac{\alpha \pi}{2} \pm i (q-p) \sin \frac{\alpha \pi}{2} \right)
  \;\;\;\; (\alpha \neq 1),
  \label{Eq:Levy}
\end{equation}
where the upper sign applies when $\zeta > 0$, and the lower for
$\zeta < 0$.
This is a well-known form of the L\'evy stable characteristic
function, and we obtain a L\'evy stable distribution $P_{L}(x)$
through an inverse Fourier transform (see Figs.~\ref{Fig:01} and
\ref{Fig:02}).

This characteristic function satisfies $n \psi_L(\zeta) =
\psi_L(n^{1/\alpha} \zeta)$, which means that the corresponding
probability distribution is stable to convolution, i.e.,
\begin{equation}
  P_{L}^{n}(x) = n^{-1/{\alpha}} P_{L}^{1}(n^{-1/{\alpha}} x).
\end{equation}
Here $P^{n}(x)$ denotes a $n$-times convoluted distribution of
$P^{1}(x) \equiv P(x)$, i.e., $P^{n}(x) = (2 \pi)^{-1} \int e^{n
  \psi(\zeta)} e^{-i \zeta x} d\zeta$.
The L\'evy stable distribution $P_{L}(x)$ is symmetric when $q-p = 0$,
and one-sided when $q-p = \pm 1$ and $0 < \alpha < 1$.\footnote{The
  distribution is one-sided to the right ($x>0$) when $q-p = -1$ and
  to the left ($x<0$) when $q-p = 1$.}
It has a power-law tail of the form $|x|^{-1-\alpha}$, and the
absolute moment $\langle x^q \rangle := \int_{-\infty}^{\infty} |x|^q
P_{L}(x) dx$ does not exist for $q \geq \alpha$.

Now, let us truncate this L\'evy stable distribution following
Koponen\cite{Koponen}. We introduce a cut-off parameter $\lambda > 0$
and truncate the original $f_{L}(x)$ in Eq.~(\ref{Eq:increments}) as
\begin{equation}
  f_{TL}(x) = \left\{
    \begin{array}{ll}
      C q \ |x|^{-1-\alpha} e^{-\lambda |x|} \;\;&\;\; (x<0), \cr
      C p \ x^{-1-\alpha} e^{-\lambda x} \;\;&\;\; (x>0). \cr
    \end{array}
  \right.
  \label{Eq:increments_cutoff}
\end{equation}
For the case $0 < \alpha < 1$, $\tau(x)$ can be omitted, and we obtain
by integration
\begin{equation}
  \psi_{TL}(\zeta ; \lambda) = C \Gamma(-\alpha) \left\{ q (\lambda +
    i \zeta)^{\alpha} + p (\lambda - i \zeta)^{\alpha} -
    \lambda^{\alpha} \right\},
  \label{Eq:truncated_Levy}
\end{equation}
or, by expanding the first two terms
\begin{equation}
  \psi_{TL}(\zeta ; \lambda) = C \Gamma(-\alpha) \left\{ \left(
      \lambda^2 + \zeta^2 \right)^{\alpha / 2} \left( \cos \alpha
      \theta + i (q-p) \sin \alpha \theta \right) - \lambda^{\alpha}
  \right\},
\end{equation}
where $\theta = \arctan \left( \zeta / \lambda \right)$.
Apart from the scale constant, this is the characteristic function of
truncated L\'evy distribution given by Koponen.\footnote{Note the
  misprint of Eq.~(3) in Ref.~\cite{Koponen}. It should read $\ln
  {\hat P}(k) = c \left\{c_0 - (k^2+{\nu}^2)^{\nu/2} / \cos(\pi \nu /
    2) ...  \right\}$.}
For the case $1 < \alpha < 2$, we use a centering function $\tau(x) =
x$ and obtain
\begin{equation}
  \psi_{TL}(\zeta ; \lambda) = C \Gamma(-\alpha) \left\{ q (\lambda +
    i \zeta)^{\alpha} + p (\lambda - i \zeta)^{\alpha} -
    \lambda^{\alpha} - i \alpha \lambda^{\alpha-1} (q-p) \zeta
  \right\}.
  \label{Eq:truncated_Levy_2}
\end{equation}
In this case, an extra term appears in addition to
Eq.~(\ref{Eq:truncated_Levy}), which induces a drift of the
probability distribution when $q-p \neq 0$.
It can easily be seen that in the limit $\lambda \to +0$, these
characteristic functions go back to the L\'evy stable characteristic
function Eq.~(\ref{Eq:Levy}).

We obtain a truncated L\'evy distribution $P_{TL}(x ; \lambda)$
through an inverse Fourier transform from
Eq.~(\ref{Eq:truncated_Levy}) or Eq.~(\ref{Eq:truncated_Levy_2}).
The parameter $\lambda$ modifies the behavior of $\psi_{TL}(\zeta ;
\lambda)$ when $\zeta$ is comparable to $\lambda$ and removes the
singularity of the characteristic function at the origin.
It thus changes the behavior of $P_{TL}(x ; \lambda)$ when $x \sim
\lambda^{-1}$ and introduces an exponential cut-off to the power-law
tail of $P_{TL}(x ; \lambda)$.
Therefore, all absolute moments $\langle x^q \rangle$ of $P_{TL}(x ;
\lambda)$ are finite in contrast to the case of the L\'evy stable
distribution $P_{L}(x)$.

The characteristic function $\psi_{TL}(\zeta ; \lambda)$ is infinitely
divisible but no longer stable. The convolutions of its corresponding
probability distribution cannot be collapsed by scaling and shifting
the variable $x$.
However, they can still be collapsed by scaling both $x$ and $\lambda$
appropriately; the characteristic function $\psi_{TL}(\zeta ;
\lambda)$ satisfies $ n \psi_{TL}(\zeta ; \lambda) =
\psi_{TL}(n^{1/{\alpha}} \zeta ; n^{1/{\alpha}} \lambda) $, which
means that the corresponding $n$-times convoluted probability
distribution satisfies
\begin{equation}
  P_{TL}^{n}(x ; \lambda) = n^{-1/{\alpha}} P_{TL}^{1} (
  n^{-1/{\alpha}} x ; n^{1/{\alpha}} \lambda).
  \label{Eq:collapse}
\end{equation}
We utilize this fact for calculating the scaling exponents of the
truncated L\'evy flights.\footnote{In the context of random processes
  we discuss below, this operation can be considered as a
  renormalization-group transformation. The limiting L\'evy stable
  process corresponds to a fixed point, and the exponent $1/\alpha$
  can be regarded as a sort of critical exponent.}

In Figs.~\ref{Fig:01} and \ref{Fig:02}, truncated L\'evy distributions
$P_{TL}(x ; \lambda)$ and their convolutions are displayed for
symmetric ($q-p = 0$) and one-sided ($q-p = -1$) cases in comparison
with the corresponding ordinary L\'evy stable distributions $P_{L}(x)$
in a log-log scale.
As expected, each truncated L\'evy distribution has a cut-off at $x
\sim \lambda^{-1} = 10^3$ after a power-law decay. The cut-off
position gradually approaches the origin with the convolution, and the
self-similarity of the convoluted distribution is broken in the tail
part.

We can extend the power-law decaying part arbitrarily longer by making
$\lambda$ smaller.
Hereafter, we assume the cut-off parameter $\lambda$ to be very small,
i.e., the cut-off is far away from the origin, since we are interested
in the transient anomalous behavior of the corresponding random
process before it converges to a Gaussian due to the central limit
theorem.

\section{Truncated L\'evy flights}

The truncated L\'evy flight\cite{Mantegna,Koponen,Dubrulle} is a
temporally discrete stochastic process characterized by the truncated
L\'evy distribution $P_{TL}(x ; \lambda)$.
At each time step $i$, a particle jumps a random distance $x(i)$
chosen independently from $P_{TL}(x ; \lambda)$.
The position $y(i)$ of the particle started from the origin is given
by $y(i) = \sum_{j=1}^{i} x(j)$.
Here we consider two representative cases of truncated L\'evy flights,
i.e., the symmetric case ($q-p = 0$, $0<\alpha<2$) and the one-sided
case ($q-p = -1$, $0<\alpha<1$).

\begin{figure}[htbp]
  \begin{center}
    \leavevmode
    \epsfxsize=0.5\textwidth
    \epsfbox{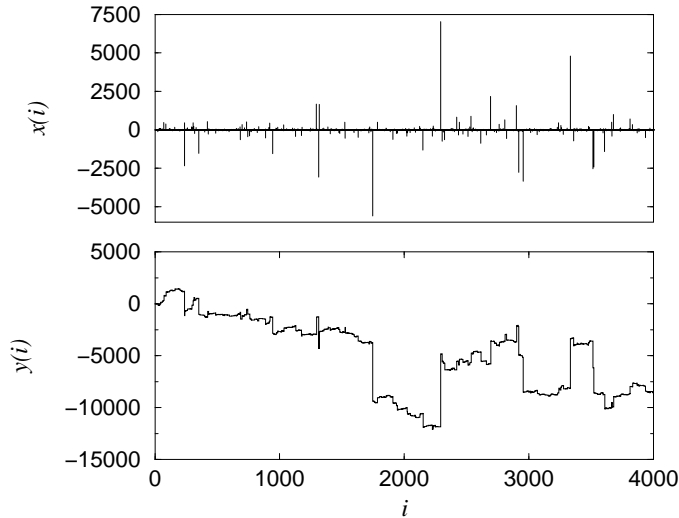}
    \caption{Typical realization of random sequence $x(i)$ (top) and 
      corresponding truncated L\'evy flights $y(i)$ (bottom) for
      $\lambda=0.001$, $\alpha=0.75$, and $q-p=0$ (symmetric case).}
    \label{Fig:03}
  \end{center}
\end{figure}

Figure~\ref{Fig:03} shows typical realizations of the jump $x(i)$ and
the position of the particle $y(i)$ for the symmetric case.
The time sequence of the jump $x(i)$ is intermittent; $x(i)$ mostly
takes small values but sometimes takes very large values.
Correspondingly, the movement of the particle $y(i)$ is also
intermittent.
This intermittency gives rise to the anomalous scaling property of the
trace of $y(i)$ in which we are interested.

\begin{figure}[htbp]
  \begin{center}
    \leavevmode
    \epsfxsize=0.5\textwidth
    \epsfbox{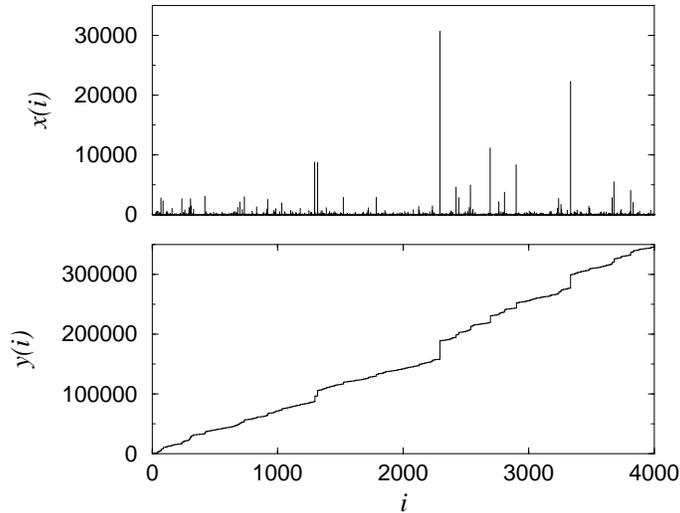}
    \caption{Typical realization of random sequence $x(i)$ (top) and
      corresponding truncated L\'evy flights $y(i)$ (bottom) for
      $\lambda=0.001$, $\alpha=0.75$, and $q-p=-1$ (one-sided case).}
    \label{Fig:04}
  \end{center}
\end{figure}

Figure~\ref{Fig:04} displays typical realizations of $x(i)$ and $y(i)$
for the one-sided case. Since the probability distribution $P_{TL}(x ;
\lambda)$ vanishes for $x < 0$, each jump takes a positive value and
the position of the particle increases monotonically. As in the
symmetric case, their time sequences are intermittent.

\section{Multi-scaling properties}

In order to characterize the intermittent time sequences shown in
Figs.~\ref{Fig:03} and \ref{Fig:04}, we introduce multi-scaling
analysis. It has been employed successfully in characterizing velocity
and energy dissipation fields of fluid turbulence and rough interfaces
of surface growth phenomena\cite{Frisch,Bohr}.

Multi-scaling analysis concerns a partition function of the measure
defined suitably for the field under consideration.
Here we focus on the apparent similarity of the intermittent time
sequences of truncated L\'evy flights to those of fluid turbulence,
i.e., the similarity of $y(i)$ in Fig.~\ref{Fig:03} to the velocity
field, and that of $x(i)$ in Fig.~\ref{Fig:04} to the energy
dissipation field.

Thus, we apply the ``multi-affine'' analysis to the trace of $y(i)$
for the symmetric sequence. The measure $h(n)$ is defined as the
absolute hight difference of $y(i)$ between two points separated by a
distance $n$, i.e., $h(n) = |y(i+n) - y(i)|$. The distribution of
$h(n)$ does not depend on $i$, since the increment of this process is
statistically stationary.
The partition function is then defined as $ Z_h(n ; q) = \langle
h(n)^{q} \rangle = \langle \left| \sum_{j=i+1}^{i+n} x(j) \right|^q
\rangle $, where $\langle ...  \rangle$ denotes a statistical average.
This function is called ``structure function'' in the context of fluid
turbulence.

On the other hand, for the one-sided case, we focus on the trace of
$x(i)$ and apply the ``multi-fractal'' analysis. The measure $m(n)$ is
defined as the area below the trace of $x(i)$ between two points
separated by a distance $n$, i.e., $m(n) = \sum_{j=i+1}^{i+n} x(j)$,
and the partition function is defined as $ Z_m(n ; q) = \langle m(n)^q
\rangle = \langle \left( \sum_{j=i+1}^{i+n} x(j) \right)^q \rangle $.

These partition functions are expected to scale with $n$ as $Z_h(n ;
q) \sim n^{\zeta(q)}$ and $Z_m(n ; q) \sim n^{\tau(q)}$ for small $n$.
Further, if these scaling exponents $\zeta(q)$ and $\tau(q)$ exhibit
nonlinear dependence on $q$, the corresponding measures $h(n)$ and
$m(n)$ are called multi-affine and multi-fractal, respectively.
\footnote{The multi-fractal partition function $Z_m(n ; q)$ defined
  here is different from the traditional one that is defined as
  $N(n)^{-1} \langle m(n)^q \rangle$, where $N(n)$ is a number of
  boxes of size $n$ that are needed to cover the whole sequence. This
  makes a difference of $-1$ to the scaling exponent $\tau(q)$ for the
  one-dimensional case considered here.}

\begin{figure}[htbp]
  \begin{center}
    \leavevmode
    \epsfxsize=0.5\textwidth
    \epsfbox{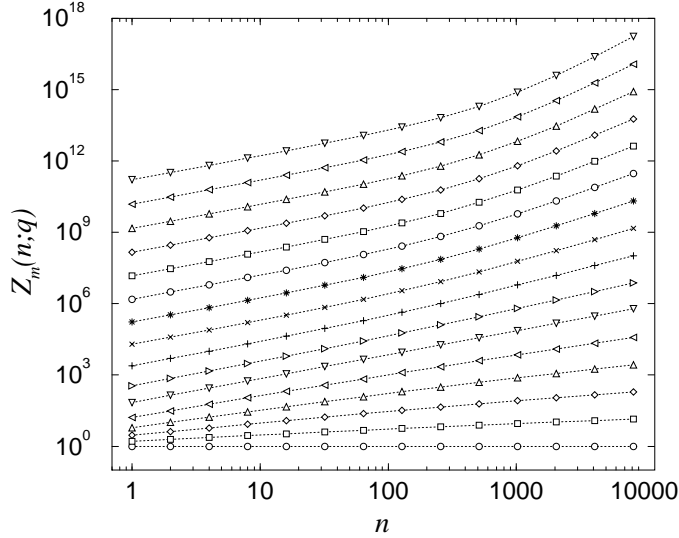}
    \caption{Partition functions $Z_m(n ; q)$ for $q=0.0$ - $3.0$
      at intervals of $0.2$. The bottom line corresponds to $q=0.0$,
      and the top one to $q=3.0$.}
    \label{Fig:05}
  \end{center}
\end{figure}

\begin{figure}[htbp]
  \begin{center}
    \leavevmode
    \epsfxsize=0.5\textwidth
    \epsfbox{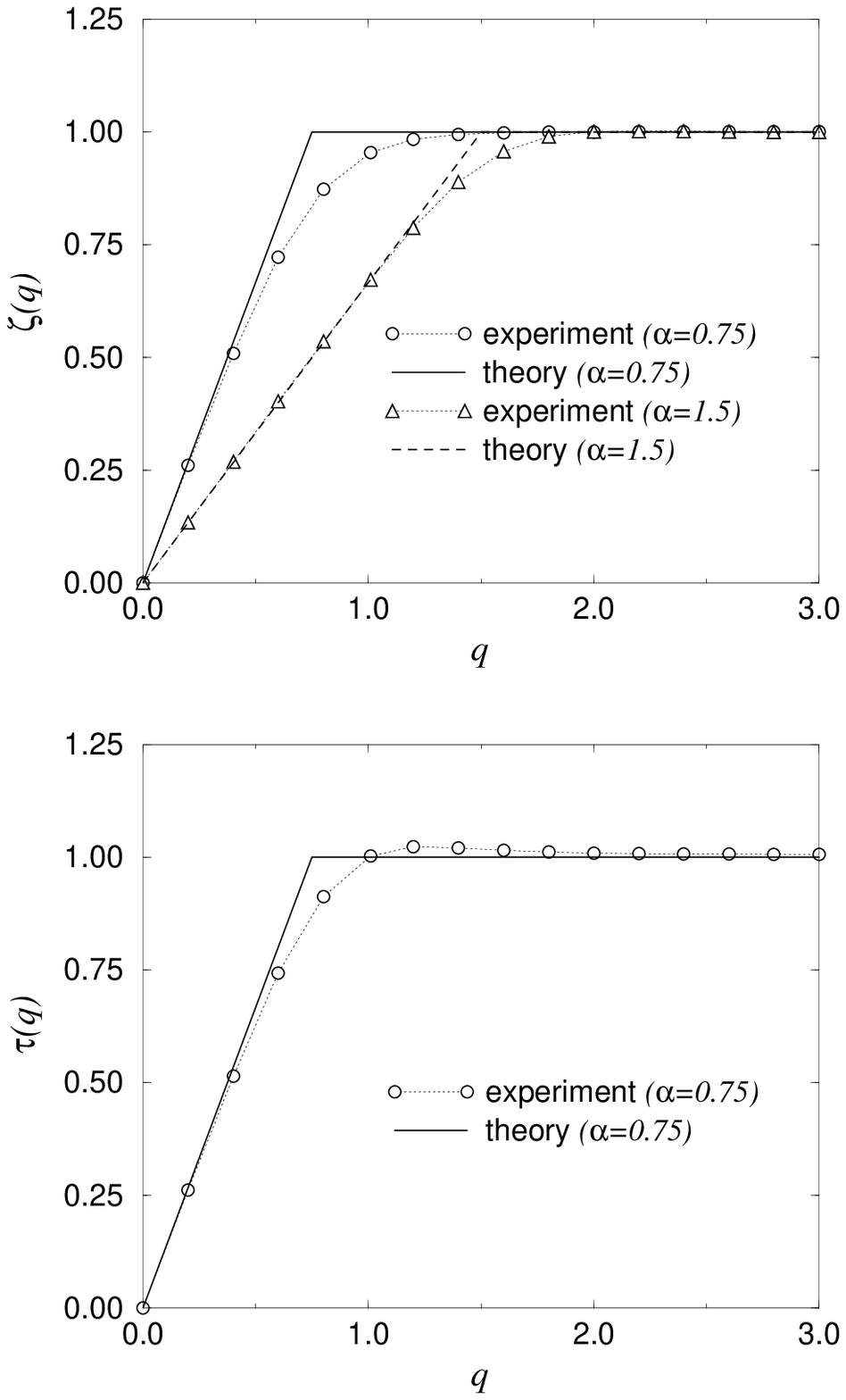}
    \caption{Multi-scaling exponents $\zeta(q)$ and $\tau(q)$
      for the symmetric (top) and one-sided (bottom) cases. The
      cut-off parameter $\lambda$ is set at $0.00001$.}
    \label{Fig:06}
  \end{center}
\end{figure}

In Fig.~\ref{Fig:05}, we display the partition functions $Z_m(n ; q)$
for several values of $q$ for the one-sided case in a log-log
scale. As can be seen from the figure, each partition function
exhibits power-law dependence on $n$ for small $n$. We obtain a
similar figure for the partition functions $Z_h(n ; q)$ for the
symmetric case.
The corresponding scaling exponents $\zeta(q)$ and $\tau(q)$ are shown
in Fig.~\ref{Fig:06}. Each curve exhibits strong non-linear dependence
on $q$; it is linear for small $q$ and constant for large $q$. Thus,
the sequences generated by truncated L\'evy flights clearly possess
multi-scaling properties.

\section{Derivation of the scaling exponents}

Now let us derive the scaling exponents $\zeta(q)$ and $\tau(q)$ from
the characteristic function Eq.~(\ref{Eq:truncated_Levy}). It is clear
from the definition of the partition functions that they can be
calculated once we know the probability distribution of the sum of
random variables $z(n) := \sum_{j=1}^{n} x(j)$.
Since the jumps are independent from each other, the probability
distribution of $z(n)$ is given by a $n$-times convolution of the
truncated L\'evy distribution $P_{TL}(x ; \lambda)$, i.e., by
$P_{TL}^{n}(x ; \lambda)$.

As we explained previously, $P_{TL}^{n}(x ; \lambda)$ can easily be
obtained from the original $P_{TL}(x ; \lambda)$ by scaling $x$ and
$\lambda$.
Making use of this fact, the $q$-th absolute moment $\langle z(n)^q
\rangle$ of $z(n)$ can be calculated as
\begin{eqnarray}
  \langle z(n)^q \rangle_{\lambda}
  &=& A \int_{0}^{\infty} z^q P_{TL}^{n}(z ; \lambda) dz
  = A \int_{0}^{\infty} z^q n^{-1/{\alpha}}
  P_{TL}(n^{-1/{\alpha}} z ; n^{1/{\alpha}} \lambda) dz \cr \cr
  &=& n^{q / \alpha} A \int_{0}^{\infty} z^q P_{TL}(z ; n^{1/{\alpha}} \lambda) dz
  = n^{q / \alpha} \langle z(1)^q \rangle_{n^{1/{\alpha}} \lambda},
\end{eqnarray}
where the constant $A$ is $2$ for the symmetric case and $1$ for the
one-sided case. Here we explicitly indicated the parameter $\lambda$
of the distribution $P_{TL}(x ; \lambda)$ as the subscript of the
average.
Note that if $\langle z(1)^q \rangle_{\lambda}$ does not depend on
$\lambda$, the scaling exponent is simply given by a linear function
$q / \alpha$, and the process does not exhibit multi-scaling. This is
the case for the ordinary L\'evy stable distribution.

Thus, all we need to calculate is the moment $\langle z(1)^q
\rangle_{\lambda}$. However, of course, an analytical expression for
$P_{TL}(x ; \lambda)$ is not attainable except a few specific cases.
Here we adopt an approximation which utilizes the facts that the
truncated L\'evy distribution is different from the ordinary L\'evy
stable distribution only in the tail part, and that it has a power-law
decaying part with an exponent $-1-\alpha$ in the middle (see
Figs.~\ref{Fig:01} and \ref{Fig:02}).
Therefore, it is expected that the moment whose degree $q$ is lower
than $\alpha$ is almost the same as that obtained from the ordinary
L\'evy stable distribution, and the moment for $q > \alpha$ is mostly
determined by the asymptotic tail of the truncated L\'evy
distribution. Note that the moment for $q > \alpha$ does not exist for
the ordinary L\'evy stable distribution.

(i) Lower moments ($0 < q < \alpha$).

We can approximate $P_{TL}(x ; \lambda)$ by $P_{L}(x)$ in this case,
since they are different only in their tail parts, which does not
contribute to the moments lower than $\alpha$ significantly.
Therefore, $\langle z(1)^q \rangle_{\lambda}$ does not depend on
$\lambda$ approximately, and we obtain $\langle z(n)^q \rangle \simeq
const. \ n^{q / {\alpha}}$. Thus, the scaling exponents $\zeta(q)$ and
$\tau(q)$ are given by $q / {\alpha}$ for $0 < q < \alpha$.
The broken self-similarity of the distribution is not important in
this regime.

(ii) Higher moments ($q > \alpha$).

Since the tail of the distribution mainly contributes to the higher
moments, we can calculate $\langle z(1)^q \rangle_{\lambda}$
approximately if we know the asymptotic form of the distribution.
For this purpose, we expand the characteristic function as in the case
of the series expansion of ordinary L\'evy stable
distribution\cite{Feller}.
By expanding the integrand, the truncated L\'evy distribution can be
expressed as
\begin{eqnarray}
  P_{TL}(x ; \lambda) &=& e^{-C \Gamma(-\alpha) \lambda^{\alpha}}
  \frac{1}{2 \pi} \int_{-\infty}^{\infty} \exp \left[ C
    \Gamma(-\alpha) \left\{ q(\lambda + i \zeta)^{\alpha} + p(\lambda -
      i \zeta)^{\alpha} \right\} \right] e^{-i \zeta x} d\zeta \cr &=&
  Re \frac{-i}{\pi x} e^{-C \Gamma(-\alpha) \lambda^{\alpha}}
  \sum_{k=0}^{\infty} \frac{1}{k!} \left\{ C \Gamma(-\alpha)
    \lambda^{\alpha} \right\}^{k} \int_0^{\infty} \left[ q\left(
      1+\frac{z}{\lambda x} \right)^{\alpha} + p\left(
      1-\frac{z}{\lambda x} \right)^{\alpha} \right]^{k} e^{-z} dz
  \;\;\;\;\;\;\;\;
  \label{Eq:expansion}
\end{eqnarray}
for the case $0 < \alpha < 1$.
At the lowest order ($k=1$), we recover the original distribution of
the increments:
\begin{equation}
  P_{TL}(x ; \lambda) \sim e^{-C \Gamma(-\alpha) \lambda^{\alpha}} C p
  x^{-1-\alpha} e^{-\lambda x} \;\;\;\; (x \gg 1).
\end{equation}
With this approximation, the moment $\langle z(1)^{q}
\rangle_{\lambda}$ is calculated as
\begin{equation}
  \langle z(1)^{q} \rangle_{\lambda} \simeq e^{-C \Gamma(-\alpha)
    \lambda^{\alpha}} C p \int_{0}^{\infty} z^{q-1-\alpha} e^{-\lambda
    z} dz = e^{-C \Gamma(-\alpha) \lambda^{\alpha}} C p \Gamma(1+q)
  \lambda^{-q+\alpha},
\end{equation}
and we obtain
\begin{equation}
  \langle z(n)^{q} \rangle_{\lambda} = n^{q / \alpha} \langle z(1)^{q}
  \rangle_{n^{1/{\alpha}} \lambda} \sim n^{q / \alpha} \left (
    n^{1/{\alpha}} \lambda \right)^{-q+\alpha} \sim n^{1}.
\end{equation}
Thus, the scaling exponents $\zeta(q)$ and $\tau(q)$ are given by $1$
for $q > \alpha$.

In summary, we approximately derived the following expression for the
multi-scaling exponents $\zeta(q)$ and $\tau(q)$:
\begin{equation}
  \zeta(q), \tau(q) = \left\{
    \begin{array}{cc}
      q/{\alpha} & \;\;\;\; (0 < q < \alpha), \cr \cr
      1 & \;\;\;\; (q > \alpha).
    \end{array}
  \right.
  \label{Eq:bifractal}
\end{equation}
Note that the above approximation becomes more and more accurate as we
extend the power-law decaying part by decreasing $\lambda$, and this
result is exact in the asymptotic limit.
In Fig.~\ref{Fig:06}, these theoretical curves are compared with the
experimental results. Except for small deviations near the transition
points, the theoretical results well reproduce the experimental
results.
Although our estimation here is done for the case $0 < \alpha < 1$,
the theoretical result Eq.~(\ref{Eq:bifractal}) seems to be also
applicable for $1 < \alpha < 2$.\footnote{This implies that
  Eq.~(\ref{Eq:expansion}) still has its meaning as an asymptotic
  expansion for $1 < \alpha < 2$. This is proved for the case of
  ordinary L\'evy stable distribution\cite{Bergstrom}.}
This type of simple multi-scaling is sometimes called
``bi-fractality'' and is known, for example, in randomly forced
Burgers' equation\cite{Frisch}.

\section{Discussion}

We analyzed the multi-scaling properties of the truncated L\'evy
flights based on the smooth truncation introduced by Koponen.
As Dubrulle and Laval claimed, truncation is essential for the
multi-scaling properties to appear.
We clarified this fact and derived the functional form of the scaling
exponents for both symmetric and one-sided cases.

As we mentioned previously, the cut-off parameter $\lambda$ may
represent the finiteness of the system under consideration. Then it
would be natural to assume $\lambda$ as a decreasing function of the
system size $L$, for example $\lambda = L^{-1}$, and we can think of
the system-size dependence of the truncated L\'evy flight. Of course,
distribution functions $P_{TL}$ for different system sizes cannot be
collapsed simply by rescaling $x$, i.e., finite-size scaling does not
hold.\footnote{We may also take a different viewpoint, where $\lambda$
  is another tunable parameter independent of the system size
  $L$. Then Eq.~(\ref{Eq:collapse}) may be viewed as a finite-size
  scaling relation between systems of size $L/n$ and $L$, similar to
  that in statistical mechanics.}
However, approximate finite-size scaling relations still hold
separately, if we divide $P_{TL}$ into two parts at the power-law
decaying part, i.e., into the self-similar part and the tail part.
As we explained, the self-similar part is insensitive to $\lambda =
L^{-1}$, and $P_{TL}$ for different system sizes collapse to a single
curve in this region without rescaling.
On the other hand, the tail part is asymptotically given by $P_{TL}(x ;
L^{-1}) \sim x^{-1-\alpha} e^{-x/L}$, which can be rescaled as
$L^{1+\alpha} P_{TL}(x L ; L^{-1})$ to give a universal curve.
Thus, we have two different approximate finite-size scaling relations
in different regions of $x$, which is separated by the power-law
decaying part.
Of course, this is closely related to the bi-fractal behavior of the
scaling exponents.
Similar asymptotic finite-size scaling is also reported in the
sandpile models of self-organized criticality\cite{Chessa}.

Recently, Chechkin and Gonchar\cite{Chechkin} discussed the finite
sample number effect on the scaling properties of a stable L\'evy
process. (They treated only the symmetric case using a different
argument from ours, which was more qualitative and therefore more
general in a sense.)
They claimed that, due to the finite sample number effect,
``spurious'' multi-affinity is observed in the numerical simulation
and derived the ``spurious'' multi-scaling exponent.
Interestingly, or in some sense obviously, their ``spurious''
multi-scaling exponent is the same as our multi-scaling exponent,
since the truncation of the power-law by $\lambda$ can also be
interpreted as mimicking the finite sample number effect of
experiments.

Our calculation presented in this paper is similar to our previous
work\cite{Nakao} on the multi-scaling properties of the amplitude and
difference fields of anomalous spatio-temporal chaos found in systems
of non-locally coupled oscillators.
The distribution treated there was not the truncated L\'evy type but
had a form like $(1+x^2)^{\alpha/2} e^{-\lambda |x|}$. Since this form
is easily generated by a simple multiplicative stochastic process,
some attempts have been made to model the economic activity using this
type of distribution\cite{Sato,Sornette}.

The bi-fractal behavior of the scaling exponent is the simplest case
of multi-scaling, while experimentally observed scaling exponents
usually exhibit more complex behavior.
Actually, it has long been discussed in the context of fluid
turbulence what the shape of the distribution should be to reproduce
the experimentally observed scaling exponent\cite{Frisch,Bohr,Benzi}.
In order to reproduce the behavior of the scaling exponent more
realistically in the framework of the truncated L\'evy flights
discussed here, introduction of correlations to the random variables
will be necessary. Studies in this direction are expected.

\acknowledgments
  The author gratefully acknowledges Professor Michio Yamada and
  University of Tokyo for warm hospitality. He also thanks
  Dr. A. Lema\^{\i}tre, Dr. H. Chat\'e, and anonymous referees for
  useful comments. This work is supported by the JSPS Research
  Fellowships for Young Scientists.

\end{document}